\documentclass[prl,twocolumn,aps,superscriptaddress,longbibliography]{revtex4-1}
\usepackage{graphicx,color,setspace}

\usepackage{xcolor}
\usepackage{bm}

\usepackage[normalem]{ulem}
\begin{document}

\title{Towards glasses with permanent stability}

\date{February 14, 2021}

\author{Taiki Yanagishima}
\affiliation{Department of Chemistry, Physical and Theoretical Chemistry Laboratory, University of Oxford, South Parks Road, OX1 3QZ, United Kingdom}
\affiliation{Department of Physics, Graduate School of Science, Kyoto University, Kitashirakawa Oiwake-cho, Sakyo-ku, Kyoto, 606-8502, Japan}
\author{John Russo}
\affiliation{Department of Physics, Sapienza University of Rome, P. le Aldo Moro 5, 00185 Rome, Italy}
\author{Roel P. A. Dullens}
\affiliation{Department of Chemistry, Physical and Theoretical Chemistry Laboratory, University of Oxford, South Parks Road, OX1 3QZ, United Kingdom}
\affiliation{Institute for Molecules and Materials, Radboud University, Heyendaalseweg 135, 6525 AJ Nijmegen, The Netherlands}
\author{Hajime Tanaka}
\email{tanaka@iis.u-tokyo.ac.jp}
\affiliation{Department of Fundamental Engineering, Institute of Industrial Science, University of Tokyo, 4-6-1 Komaba, Meguro-ku, Tokyo 153-8505, Japan}
\affiliation{Research Center for Advanced Science and Technology, University of Tokyo, 4-6-1 Komaba, Meguro-ku, Tokyo 153-8505, Japan}

\begin{abstract}
Unlike crystals, glasses age or devitrify over time, reflecting their non-equilibrium nature. This lack of stability is a serious issue in many industrial applications. Here, we show by numerical simulations that the devitrification of quasi-hard-sphere glasses is prevented by suppressing volume-fraction inhomogeneities. A monodisperse glass known to devitrify with `avalanche'-like intermittent dynamics is subjected to small iterative adjustments to particle sizes to make the local volume fractions spatially uniform. We find that this entirely prevents structural relaxation and devitrification over aging time scales, even in the presence of crystallites. There is a dramatic homogenization in the number of load-bearing nearest neighbors each particle has, indicating that ultra-stable glasses may be formed via `mechanical homogenization'. Our finding provides a physical principle for glass stabilization and opens a novel route to the formation of mechanically stabilized glasses.
\end{abstract}

\maketitle

Glassy materials are known to spontaneously age and devitrify~\cite{angell2000relaxation}. The gradual aging-driven drift in the physical properties of glasses over time is a serious issue for their applications. For example, localized crystallization can be detrimental to aqueous media in cryogenics \cite{Han2007,Hunt2011,Sansinena2014}, pharmaceuticals \cite{craig1999relevance,Knapik-Kowalczuk2018,Salunkhe2019} and optical media \cite{Vestel2003,Lin2019,Ordu2020}. There are also some instances when devitrification may be empirically tailored to tune the properties of amorphous phases, such as in metallic glasses \cite{Greer2001,Fornell2010,Louzguine-Luzgin2014}. However, despite its ubiquity in supercooled systems, the exact mechanism by which aging and devitrification occur is not yet understood. Recent simulations have found that unique, intermittent dynamics can be seen in poor glass formers at deep supercooling, accompanied by collective `avalanche'-like displacements \cite{Sanz2014}. Further analysis of these `avalanche' events found a statistical correlation between avalanche incidence and low local volume fraction \cite{Yanagishima2017}. Experimental work on colloidal glasses has also corroborated a relationship between local structure, local dynamics, and subsequent devitrification  \cite{Simeonova2006,Ganapathi2020}. However, a physical \emph{mechanism} for devitrification has not yet been presented. In particular, the relationship between mechanical stability and devitrification or aging in these systems as introduced in Ref.~\cite{Yanagishima2017} remains to be addressed.

Given the correlation between avalanche initiation and local volume fraction~\cite{Yanagishima2017}, we take an orthogonal approach to previous works by \emph{removing} inhomogeneities in local volume fractions and studying the subsequent dynamics. This strategy is closely linked to studies of hyperuniform packings, themselves characterized by suppressed long-range spatial correlations in local density~\cite{Torquato2018,Kim2019}. The system we study is a dense glass of particles interacting via a repulsive Weeks-Chandler-Andersen (WCA) potential at thermal energy $k_{\rm B}T=0.025$, propagated over time using standard Brownian Dynamics (overdamped Langevin dynamics) \cite{Ermak1978}. The system is commonly regarded as closely approximating hard spheres or experimental colloidal systems and has been widely used as an ideal model system to study aging and devitrification~\cite{Simeonova2006,martinez2008slow,lynch2008dynamics,Zaccarelli2009a,zargar2013direct,kawasaki2014structural,Sanz2014,Yanagishima2017,Ganapathi2020}. Firstly, `conventional' glassy (CG) states were prepared using a modified version of the Lubachevsky-Stillinger algorithm \cite{Lubachevsky1990}, identical to the method used in previous work \cite{Yanagishima2017}. Initial particle sizes are monodisperse, $\sigma_i = 1$ for all $i$, where $\sigma_i$ is the size of particle $i$ in the repulsive WCA potential. After an initial configuration is generated at some bulk volume fraction $\phi_0$, the energy of the packing is minimized using the FIRE algorithm \cite{Bitzek2006} to relax residual stresses introduced by the quench.

\begin{figure}[t]
    \centering
    \includegraphics[width=7.cm]{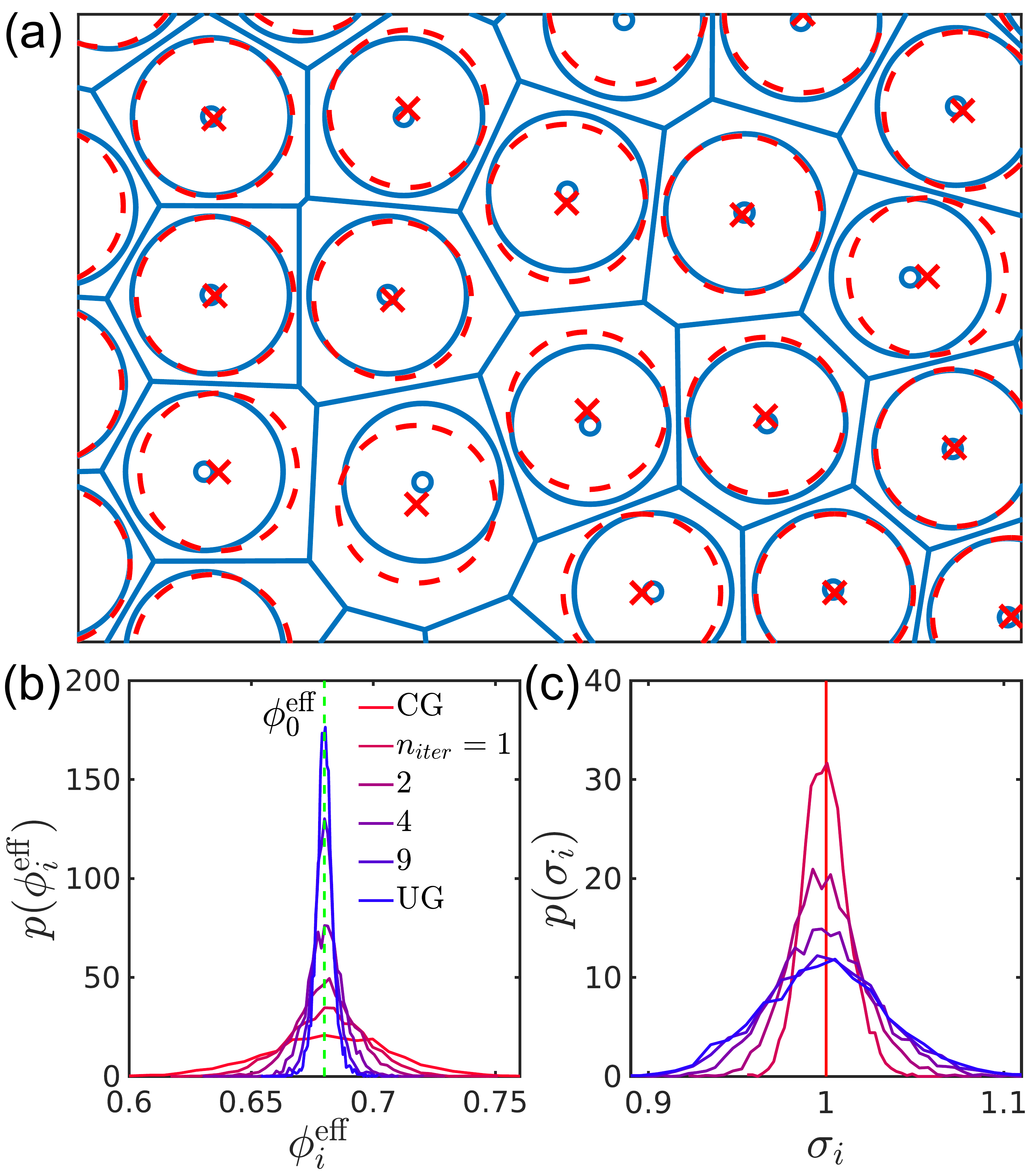}
    \caption{Homogenization of local volume fractions. (a) Illustration of the $\phi_i$ flattening algorithm. A Voronoi tessellation is taken of an energy minimized configuration (solid blue). Particles are resized to match the local $\phi_i$ to the $\langle\phi_i\rangle$ (dashed red) before being relocated to the center of the Voronoi cell (crosses). The energy is minimized again before the process is repeated until this global operation no longer reduces the standard deviation in the local volume fractions. (b) The probability density function of effective local volume fractions $\phi^{\rm eff}_i$ with each iteration. (c) The probability density function of particle sizes $\sigma_i$ with each iteration.}
    \label{fig:algo}
\end{figure}

In order to flatten the spatial volume fraction profiles, we adapt a recently developed method for making hyperuniform packings \cite{Kim2019}. The local volume fraction, $\phi_i$, of particle $i$ is estimated using a radical Voronoi tessellation. Particles are then resized such that, for each Voronoi cell, $\pi\sigma_i^3/(6v_i) = \phi_0$, where $v_i$ is the local Voronoi volume. The particle is then replaced in the center of the cell. As radical Voronoi cell boundaries change when the particle is resized, the state is no longer at an energy minimum. Thus, the packing is rerun through the FIRE algorithm, and the process repeated, checking that the standard deviation of local densities $\Delta\phi_i/\langle\phi_i\rangle$ decreases (see Fig.~S1 \cite{Supple}). This iterative process is stopped once changes to particle sizes no longer result in a narrowing of the local volume fraction distribution. We call this final glass state a ``uniform glass (UG)'', with its uniform local volume fraction over space. An illustration of the algorithm, the narrowing of the volume fraction profile, and the slight broadening of the particle size $\sigma_i$ distribution are shown in Figs.~\ref{fig:algo}(a), (b), and (c), respectively. In this paper, we express volume fractions using an effective volume fraction $\phi^{\rm eff} = \phi \times (1.0953^3)$, a mapping that matches the melting and freezing of a monodisperse WCA system at $k_{\rm B}T=0.025$ to hard spheres \cite{Heyes2006}.

\begin{figure}[b]
    \centering
    \includegraphics[width=6.5cm]{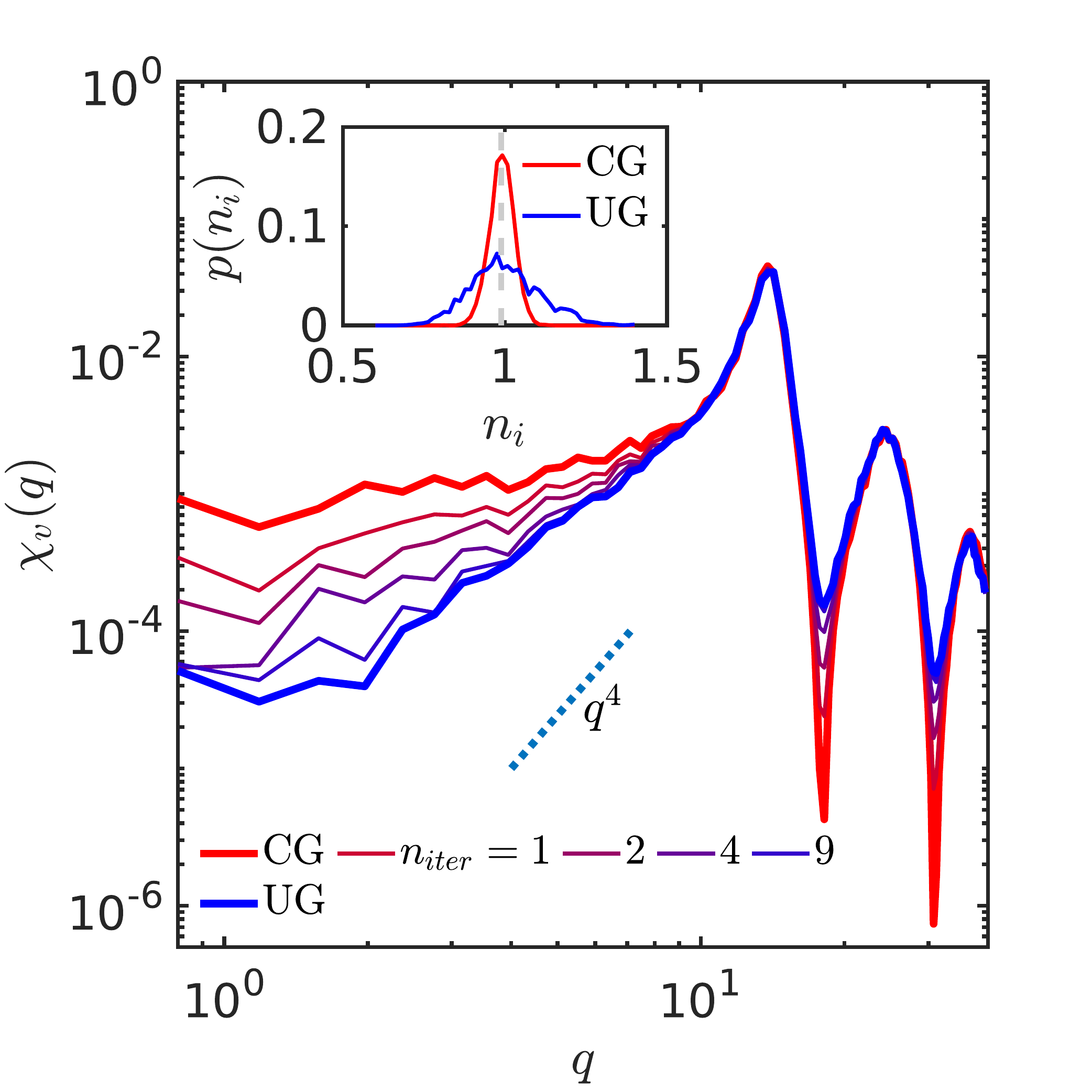}
    \caption{Spectral density of static structures, from CG to UG. Spectral densities $\chi_v$ are given for progressive iterations as the CG state is transformed into a UG state. As the iterations progress, $\lim_{q \rightarrow 0} \chi_v(q)$ systematically decreases, though the scaling does not reach the $q^4$-scaling of a class I hyperuniform state. (inset) Distribution of local number densities $n_i$. The UG state features a significantly wider distribution of $n_i$.
    }
    \label{fig:Sq}
\end{figure}

To quantify the action of this $\phi_i$ flattening algorithm, we generate 50 independent CG/UG pairs. When $\phi^{\rm eff}_0 = 0.68$, the initial CG glasses we generate have a $\phi_i$ distribution with a standard deviation $\Delta\phi_i/\langle\phi_i\rangle$ ranging from 3.7\% to 3.9\% over the 50 samples. After flattening, this is reduced to 0.37\% to 0.95\% (mean 0.63\%), with a particle size polydispersity of 3.0\% to 3.7\% (mean 3.4\%). Notably, this polydispersity is significantly lower than the dispersion required to prevent crystallization \cite{Moriguchi1993,Bartlett1997,Zaccarelli2009a,tanaka2010critical}; i.e., one would still expect a strong thermodynamic driving force towards crystallization. We emphasize that this polydispersity approaches the one present in the most monodisperse colloidal suspensions used to experimentally study crystallization~\cite{Palberg2014}. To show that the properties of the UG state are independent of polydispersity, we generate independent CG states that share the same particle size distribution of UG states, and refer to them as re-CG states in the following.

The fact that the UG states are prepared by suppressing local $\phi_i$ fluctuations suggests a natural connection between these states and hyperuniform configurations~\cite{zachary2011hyperuniform,zachary2011hyperuniformity,Torquato2018,Kim2019}, as mentioned above. To verify this, in Fig.~\ref{fig:Sq}, we plot the spectral density of the configurations $\chi_v$, defined as a power spectrum $\chi_v(q) = V^{-1} \langle I_q I^*_q \rangle$. $I_q$ is the Fourier transform of $I(r)$, the indicator function for a polydisperse suspension of hard spheres defined as $I({\bf r}) = \Sigma_i \Theta(|{\bf r - r_i}|-R_i)$, where $\bf{r}_i$ is the position of particle $i$, $R_i =  \sigma_i/2$ is the radius, $\Theta$ is the Heaviside function, and $V$ is the total volume \cite{zachary2011hyperuniform,zachary2011hyperuniformity}. In the context of an experiment, $\chi_v$ is simply the power spectrum of a binary `image' of particles. It is clear that the $q \rightarrow 0$ limit of $\chi_v$ is diminished with progressive iterations. However, it is also clear that the $q^4$-scaling at low $q$ expected for class I hyperuniform systems \cite{Torquato2018} is not reached. We also note that the distribution of number densities $n_i$ significantly broadens; since the algorithm keeps the volume fraction constant, the size polydispersity results in a broader number density distribution, as shown in the inset of Fig.~\ref{fig:Sq}. Overall, it is clear that our UG states are \emph{approaching} a hyperuniform state, but cannot be classed as hyperuniform themselves.

\begin{figure}
    \centering
    \includegraphics[width=1.0\columnwidth]{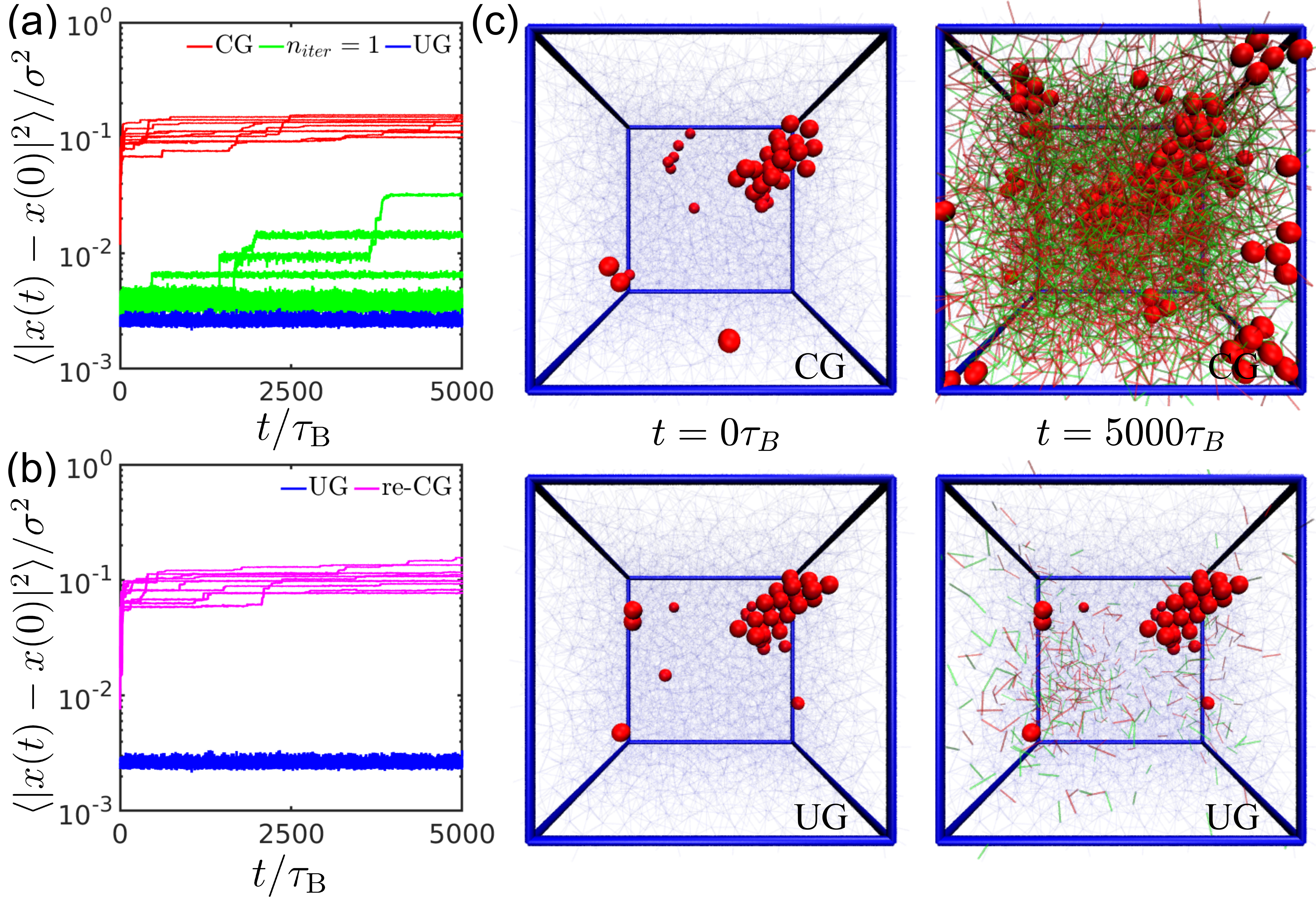}
    \caption{Impact of flattening $\phi_i$ over space on avalanche events and devitrification. (a) Squared displacements over ten 5000$\tau_B$ runs initiated from the same CG state, the state after one iteration of the $\phi_i$ flattening algorithm, and the final UG state, as indicated. (b) Squared displacements over ten 5000$\tau_B$ runs initiated from ten independently generated CG states using the same particle sizes as in the UG state (re-CG). Squared displacements for the UG state from (a) are also drawn for comparison. (c) Change in crystalline particles and force network for corresponding CG and UG states. The CG-generated state experiences growth in crystallinity, while the UG state created from it does not change. Changes in force connections between nearest neighbors are also shown. The thin blue lines indicate the original network; red and green lines at $t = 5000\tau_B$ indicate connections broken and formed.}
    \label{fig:Xt}
\end{figure}

Now, we turn our attention to the dynamics. Brownian dynamics is applied to CG states, states reached after $n_{iter}$ iterations of the algorithm, and final UG states to observe the dynamics over 5000$\tau_B$, where $\tau_B$ is the Brownian relaxation time in the ultra-dilute limit, $\tau_B = 6\pi\eta \langle\sigma\rangle^3/k_BT$, with $\eta$, the viscosity of the medium, set to 1 in the simulation. On inspecting the average squared displacement of the particles from their initial configuration, we find a drastic change in the dynamical behavior of the system. Squared displacements for 10 isoconfigurational trajectories are shown in Fig.~\ref{fig:Xt}(a). The CG states showed intermittent dynamics, the same avalanche phenomenon reported previously~\cite{Sanz2014,Yanagishima2017}. However, even a single iteration of the $\phi_i$ flattening algorithm significantly reduces both the frequency and size of the avalanches. Further iterations quickly reduce the frequency to zero, rendering the dynamics free of avalanches.

The lack of avalanche-like dynamics leads directly to a halt in the growth of pre-existing crystallites. Figure~\ref{fig:Xt}(c) shows crystalline particles in energy-minimized CG and UG states before and after dynamical propagation using Brownian dynamics. Crystallites are detected using nearest neighbor bond coherence in bond-orientational order parameter $q_6$, as in Refs.~\cite{Steinhardt1983,Russo2012a}. Note that the UG states see no significant growth, as if the volume fraction flattening effectively `freezes' post-critical nuclei. We stress that the bulk volume fraction $\phi_0$ and temperature remain unchanged. This is doubly clear by visualizing crystalline particles in CG and UG states over time as they thermally fluctuate (see Supplementary Movie~\cite{Supple}). Despite some fluctuations in bond coherence, the crystallinity of the UG state remains unchanged, while the CG state experiences significant crystal growth. This strongly corroborates previous findings that the avalanches provide the necessary perturbation for existing crystals to grow at deep supercooling~\cite{Sanz2014,Yanagishima2017}. It is worth emphasizing that this `stability' against avalanche-mediated crystallization is achieved by introducing only a \emph{minimal} polydispersity in particle size. For the UG configurations used to calculate squared displacements in Figs.~\ref{fig:Xt}(a) and (b), this is 3.60\%. Figure~\ref{fig:Xt}(b) compares the evolution of 10 UG states and their respective re-CG states (i.e., having the same size distribution but with an inhomogeneous local volume fraction distribution), showing that re-CG states are prone to avalanche displacements and that size polydispersity alone does not play a role in the stabilization of the UG states.

\begin{figure}
    \centering
    \includegraphics[width=1.0\columnwidth]{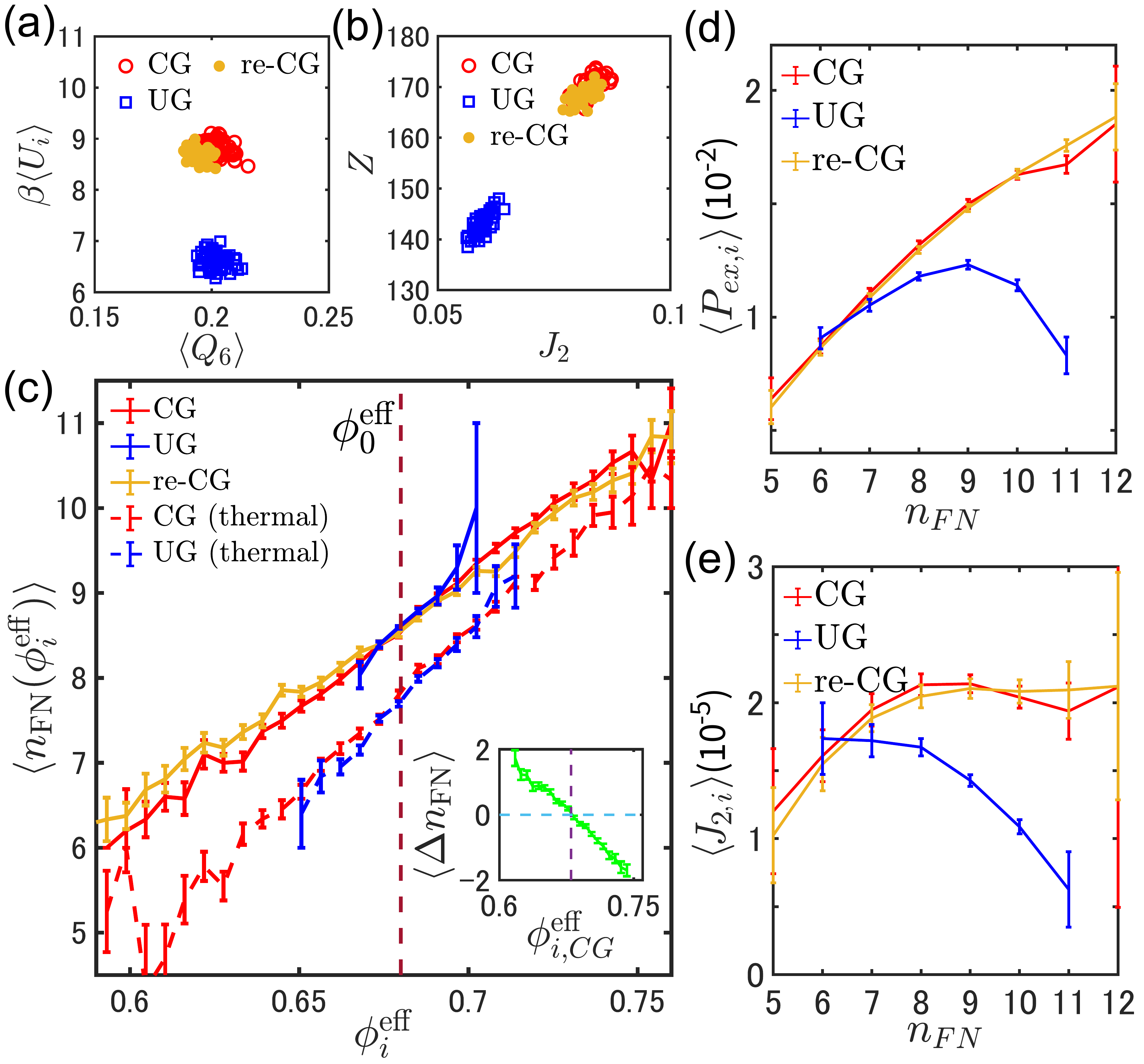}
    \caption{Impact of $\phi^{\rm eff}_i$ homogenization on structure, thermodynamics, and mechanics. (a) Average energy per particle $\beta\langle{U}\rangle$ vs. $Q_6$ for 50 independent pairs of CG/UG configurations, and re-CG states generated from the final UG particle size distribution. (b) Compressibility factor $Z = \beta P_V / \langle \sigma\rangle^3$ vs. deviatoric stress invariant $J_2$ over same states as in (a). (c) Average number of force neighbors $\langle n_{\rm FN} \rangle$ for particles with different local volume fractions $\phi^{\rm eff}_i$ in CG (red), UG (blue) and re-CG (yellow) states. Solid lines are found from energy minimized configurations, dashed lines from thermally fluctuating configurations at the beginning of the Brownian Dynamics simulation ($t = 5\tau_B$). (inset) $\langle \Delta{n}_{FN}\rangle$ as individual particles with different initial $\phi^{\rm eff}_i$ in the CG state are transformed to reach a UG state. (d) Hydrostatic contribution to pressure from individual particles $P_{ex,i}$ vs. $n_{FN}$ for CG, UG and re-CG states. (e) Deviatoric stress invariant contribution from individual particles $\langle J_{2,i} \rangle$ for CG, UG and re-CG states.}
    \label{fig:Str}
\end{figure}

We now examine the structural changes that occur going from CG to UG states. At the level of pair correlations, we see that these states are virtually identical: Figure~S2 \cite{Supple} plots the radial distribution function $g(r)$ for CG, UG, and re-CG states, showing that the $g(r)$ of UG and re-CG states (which have the same particle-size distribution) are almost indistinguishable. Pair correlations thus are unable to capture the sharp discrepancy between the dynamics of UG and re-CG states. In Fig.~\ref{fig:Str}(a), we plot the parameter $\langle{Q_6}\rangle$, a coarse-grained measure of bond-orientational order $q_6$ averaged over nearest neighbors~\cite{lechner2008accurate}, and the energy per particle $\beta \langle U_i\rangle$ for 50 independently generated CG states, corresponding UG states and re-CG states generated from the particle size distribution of the UG states. While $\langle{Q_6}\rangle$ fails to distinguish between (re-)CG and UG states, the UG states are found to be considerably lower in the energy landscape than their CG counterparts. Also the compressibility factor $Z = \beta P_V / \langle{\sigma^3}\rangle$, where $P_V$ is the virial pressure, and the deviatoric stress invariant $J_2=tr(\bm{s}^2)/2$, where $\bm{s}$ is the stress tensor, both plotted in Fig.~\ref{fig:Str}(b), show a significant drop going from (re-)CG states to UG states. Note that the drop is greater over subsequent iterations of the $\phi_i$-flattening algorithm (see Fig.~S3 \cite{Supple}).
We conclude that the system is \emph{annealed} going from CG to UG states, with drops in both internal stress and energy, but this transformation is distinct from \emph{aging}, as no significant changes in local structure occur at the pair (e.g., $g(r)$), and many-body level (e.g., $\langle{Q_6}\rangle$). Thus, the UG state is not reached thermodynamically and should not be confused with the \emph{ideal} glass, which is defined in a thermodynamic context~\cite{Cavagna2009a,tanaka2010critical,kawasaki2014structural,tanaka2019revealing}.

The central question now becomes the nature of the subtle change in structure that causes such a dramatic change in stability. To this end,  we investigate the distribution of \emph{force neighbors}, which were shown to describe the onset of mechanical stability in glasses~\cite{Yanagishima2017,tong2020emergent} and gels~\cite{tsurusawa2019direct}. For particle $i$ with size $\sigma_i$, we measure the number of nearest neighbors $j$ which are located at a distance $r$ such that $r < 2^{\frac{1}{6}}(\sigma_i + \sigma_j) \times \frac{1}{2}$, the interaction range of the WCA potential. These `force' neighbors (FN) exert a repulsive force on particle $i$, and create linkages in the force chain network of the configuration; we let $n_{\rm FN}$ be the number of force neighbors surrounding each particle. Figure~\ref{fig:Str}(c) shows $\langle n_{\rm FN} \rangle$ for particles with different local volume fraction $\phi^{\rm eff}_i$ in a typical CG (red) state and its corresponding UG (blue) state. Distributions are given for both energy minimized (i.e., inherent) states and a thermally fluctuating configuration at the beginning of a Brownian Dynamics trajectory ($t = 5\tau_B$). 

Firstly, we see that the CG state exhibits a linear relationship between the local volume fraction of a particle $\phi^{\rm eff}_i$ and the average number of force neighbors it has, $\langle n_{\rm FN}(\phi_i) \rangle$. With thermal fluctuations, the same trend is seen, except with fewer force connections. Now, going to the corresponding UG state, we find that the trend is closely preserved. The consequence of this is that the narrowing of the local volume fraction distribution directly results in a narrowing of the distribution of the number of force neighbor particles. The large error bars at the extrema are due to the small number of particles in the corresponding bins ($<10$). The narrowing of the $n_{\rm FN}$ distribution is further illustrated by considering changes in the number of force neighbors $\Delta n_{\rm FN}$ for individual particles with different $\phi^{\rm eff}_i$ in the CG state as it is transformed into a UG state (see inset of Fig.~\ref{fig:Str}(c)). This confirms that particles at lower $\phi^{\rm eff}_i$ gain force neighbors, while those with higher $\phi^{\rm eff}_i$ lose them. Both these findings indicate that the narrowing of the distribution in $\phi_i^{\rm eff}$ leads directly to a narrower, more homogeneous distribution of $n_{\rm FN}$, i.e., \emph{mechanical} homogenization. This strongly suggests that the mechanism behind the exceptional resistance to crystallization in the UG state is \emph{mechanical}. Note that the same analysis of a re-CG state recovers the same broad range of $\phi_i^{\rm eff}$ and the resulting range of $n_{FN}$ as the CG-state, corresponding to the recovery of avalanche dynamics. The same trends can be seen when CG and UG states are prepared at different volume fractions, as shown in Fig.~S4 \cite{Supple}. 

Local stresses are also affected by the homogenization of local connections. In Fig.~\ref{fig:Str} we plot both local hydrostatic (d) and deviatoric (e) pressures as a function of the number of force neighbors averaged over the same 50 independent configurations used for Figs.~\ref{fig:Str}(a) and (b). In Fig.~\ref{fig:Str}(d) we plot the average excess pressure contribution, $\langle P_{ex,i}\rangle$ for particles with different $n_{FN}$: in contrast to CG and re-CG states, whose excess pressures increase with $n_{FN}$, UG states have a maximum around $n_{FN} = 9$, close to the average $n_{FN}$ at the considered volume fraction. Similar behavior is found in the local deviatoric stress invariant $J_{2,i}$ plotted in Fig.~\ref{fig:Str}(e), in that more local contacts $n_{FN}$ results in a \emph{lower} deviatoric stress, though there is no maximum at $n_{FN} = 9$. The local arrangements of nearest neighbors in the UG state result in significant reductions in both volume (hydrostatic) and shape (deviatoric) altering stresses acting at the local level. In conclusion, the enhanced mechanical stability of the UG states is due to both a narrowing in the range of $n_{FN}$ and a reduction of local stresses.

The action of \emph{mechanical homogenization} on devitrification may be understood as rendering configurations free of force chain network defects, making the system more resistant to thermal fluctuations. The connection between mechanical rigidity and glassy dynamics is consistent with recent work, which showed that a percolated force network is spontaneously formed below an experimental glass transition temperature, leading to the emergence of shear rigidity~\cite{tsurusawa2019direct,tong2020emergent}. Applied to our system, the destabilization of this percolated force network is the only mechanism by which our configurations may rearrange, i.e., triggering avalanches and causing structural aging and devitrification~\cite{Yanagishima2017}; thus, by minimizing heterogeneities in the force network, we see a significant elevation in stability. The robustness that this imparts may be observed directly in Fig.~\ref{fig:Xt}(c) and the Supplementary Movie~\cite{Supple}, which also shows force network connections broken (red), formed (green), and kept (blue) over time. It is worth noting that the combined simulation time of the Brownian Dynamics trajectories corresponds to at least 50,000$\tau_B$. For a typical colloidal experiment using 1-micron diameter particles at room temperature in water, this corresponds to 30 hours without a \emph{single} rearrangement event, rendering our UG states `permanently' stable within the observation time. We may also demonstrate this by considering perturbations to the UG state that artificially `trigger' avalanche events in specific locations by removing or ``voiding'' single particles; due to the uniformly stable force network structure, avalanche events are highly localized, in sharp contrast to those seen for deeply supercooled conventional glasses~\cite{Sanz2014, Yanagishima2017} (see \cite{Supple} for details). All these results support the intrinsic link between the stability of uniform glasses and mechanical homogenization.

In summary, we successfully draw a correlation between the stability of a glass state and mechanical homogeneity. We believe this to be the first time avalanche dynamics has been directly related to a physical mechanism. Importantly, our results also pave the way towards developing a practical method to prepare mechanically stabilized homogeneous glasses. Experimental work~\cite{corte2008random} has shown that a colloidal system exhibits a transition to an absorbing state under oscillatory shear via so-called ``random organization'', where the critical absorbing state is, in fact, hyperuniform~\cite{hexner2015hyperuniformity,tjhung2015hyperuniform,weijs2015emergent,wilken2020hyperuniform}. Formation of a hyperuniform state in binary charged colloids has also been suggested~\cite{chen2018binary}. Given that our UG states are `on the way' to hyperuniformity, these recent studies suggest that any experimental efforts to approach such a state may readily realize mechanically stable UG states. Thus, we may realistically expect that our findings inspire experimental efforts not only to reach hyperuniformity, but to realize glasses that can resist aging. Such a method would significantly impact glass applications, as it would enable the preparation of ultra-stable glasses with unparalleled stability against degradation or devitrification over time.

We thank Eric Corwin and Jack Dale for their valuable contributions at the initial stages of this project. We are also grateful to Francesco Sciortino for many valuable suggestions and Francesco Turci for an independent review of some of the structural analyses in this paper. T.Y. and R.D. acknowledge the support of a European Research Council Grant 724834-OMCIDC. T.Y. acknowledges receipt of Grants-in-Aid for Young Scientists (JSPS KAKENHI Grant Number JP15K17734) and JSPS Research Fellows (JSPS KAKENHI Grant Number JP16J06649) from the Japan Society for the Promotion of Science (JSPS), and the Kyoto University Research Fund for Young Scientists (Start-Up) FY2021. J.R. acknowledges support from the European Research Council Grant DLV-759187. H.T. acknowledges receipt of Grants-in-Aid for Scientific Research (A) (JSPS KAKENHI Grant Number JP18H03675) and Specially Promoted Research (JSPS KAKENHI Grant Number JP20H05619) from the Japan Society for the Promotion of Science (JSPS).


%

\end{document}